\documentclass[floatfix,amssymb]{revtex4}
\usepackage[dvipsnames]{xcolor}
\usepackage{amstext}
\usepackage{amsmath}
\usepackage{latexsym}

\usepackage[english]{babel}
\usepackage[latin1]{inputenc}
\usepackage{times}
\usepackage[T1]{fontenc}
\usepackage{graphics}
\usepackage{graphicx}
\usepackage{setspace}
\usepackage{textpos}

\newcommand{\be}{\begin{equation}}
\newcommand{\ee}[1]{\label{#1} \end{equation}}
\newcommand{\ba}{\begin{eqnarray}}
\newcommand{\ea}[1]{\label{#1} \end{eqnarray}}
\newcommand{\nl}{\nonumber \\}
\newcommand{\re}[1]{(\ref{#1})}

\newcommand{\pd}[2]{ \frac{\partial #1}{\partial #2}}

\newcommand{\pv}[2]{ \frac{\delta #1}{\delta #2}}

\begin{document}
\title[From Madelung to Dilaton] 
{Splitting the source term for the Einstein equation to classical and quantum parts}
\author{T.S.~Bir\'o and P.~V\'an}
\affiliation{
  Heavy Ion Research Group\\
  MTA 
  Wigner Research Centre for Physics, Budapest 
}
\date{\today}
\begin{abstract}
We consider the special and general relativistic extensions of the action 
principle behind the Schr\"odinger equation distinguishing classical and 
quantum contributions. Postulating a particular quantum correction to the 
source term in the classical Einstein equation we identify the conformal 
content of the above action and obtain classical gravitation for massive 
particles, but with a cosmological term representing off-mass-shell 
contribution to the energy-momentum tensor. In this scenario the - on the 
Planck scale  surprisingly small - cosmological constant stems from quantum 
bound states  (gravonium) having a Bohr radius $a$ as being $\Lambda=3/a^2$. 
\end{abstract}

\maketitle
	\tableofcontents
\section{Introduction}
The enigma of the cosmological constant, in the modern view interpreted as dark 
energy, is mainly due to its surprisingly small, yet nonzero magnitude:  should 
it have namely a quantum gravity origin (analogous to a symmetric broken phase 
with nonzero Higgs fields in the Standard Model),  then the natural scale for a 
ground state (vacuum) energy density would be of $M_P^4$ order, with $M_P$ 
being the Planck mass. In the symmetric phase on the other hand it should be 
{\em exactly zero}. According to astronomical observations, however, the effect 
is about $120$ decimal orders of magnitudes too small for the energy density 
(and still $30$ orders too small in the linear energy scale) while it is 
definitely non zero  
\cite{Sit16a,Sit16a1,Sit17a1,Fri22a,Fri24a,Lem27a,Ein17a,Ein31a,Pad03a,Bur13m,
Hub29a,CarEta92a,Wei89a,Car01a,Planck13m,Pad06p}.

In spite of this naturalness problem the dark energy is responsible for 
$68-72\%$ of the evolution of the Universe observed presently in the standard 
cosmological models based on Friedmann's first calculation. A remaining 
$24-28\%$ of the effect is called dark matter, about which more ideas have been 
already discussed in the literature. The classical naturalness problem has 
probably nothing to do with quantum gravity. Its appearance may be caused by 
quantum effects on the source term in the Hilbert-Einsteinian gravity theory. 

In this paper we test a particular idea based on a conformal treatment of the  
Schr\"odinger equation: as if the  quantum mechanical problem of obtaining wave 
functions, and the special  relativistic field theory problem with scalar 
fields, could be splitted to a massive and a conformal part in line with a 
classical -- quantum partition \cite{Sch26a,Sch26a1,Sch26a2,Sch26a3}.  
Considering in a relativistic setting (but without spin effects) the free 
Klein-Gordon action is inspected. Identifying the quantum part as belonging to 
a traceless relativistic energy-momentum tensor, we suggest to generalize its 
Bohm-Takabayashi form  \cite{Boh52a,Boh52a1,Boh53a,BohVig54a,Tak52a,Tak53a} and 
connect the remaining classical part to Einstein's gravity equation 
\cite{Ein15a,Ein16a,Hil15a} in form of a dust matter source of massive point 
particles moving on Bohm trajectories. The resulting energy-momentum tensor 
from this procedure agrees with the proper  handling by variation against the 
metric tensor for the conformal invariant part of the action. In this scenario 
the quantum nature of the {scalar field} reveals itself in deviations from the 
classical on-mass-shell relation $P_\mu P^\mu=(mc)^2$, and our suggested 
natural coupling to gravity makes a simple conformal transformation of the full 
Einstein tensor expedient. After this transformation the classical part (dust 
gravity) separates from quantum effects which among others include a 
cosmological term. This term represents negative pressure e.g. for stationary 
quantum bound  states in a simple attractive $-\alpha/r$ gravitational 
potential, with a mass of about $140$ MeV for the pairwise composite object.

In this paper we first recall the Schr\"odinger equation with complex magnitude 
-- phase variables, together with the underlying action principle. Then the 
Klein-Gordon quantum action is analyzed in the same way, aiming at the 
determination of the physically correct energy-momentum tensor. Here we 
emphasize the quantum-conformal (in the Bohm-like contribution traceless) 
construction possibility. 
Based on this we apply a naturally emerging conformal transformation to the 
Einstein equation with the generalized Bohm-Takabayashi energy-momentum tensor 
as matter source. Then one identifies the classical Einstein tensor, the 
quantization volume, the conformal symmetry of the quantum part and a 
cosmological term proportional to the off-mass-shell part in the flat space.

\section{Nonrelativistic Quantum Mechanics}

Although this paper is ultimately written with the purpose of testing an idea  
about a non-conventional explanation for the origin of the classical 
cosmological constant, at this point we recapitulate a few conceptual issues in 
non-relativistic quantum mechanics because the suggested modification to the 
energy-momentum tensor, as the source term of the classical Einstein equation 
describing gravity, is motivated by a particular view on quantum binding energy 
effects. These effects will be re-interpreted in the next section in the 
framework of special relativistic quantum field theory as off-mass-shell 
contributions.

\subsection{Schr\"odinger equation with Madelung variables}

The Madelung picture \cite{Mad26a} of the Schr\"odinger equation has been 
criticized due to various reasons (see e.g. \cite{Jan69a,Boe92a,Wal94a}). On 
the other hand it is a permanent source of  inspiration equally in applied and 
fundamental quantum research. J\'anossy and his coworkers stressed the fluid 
interpretation up to its limits  \cite{Jan62a,JanZie63a,JanZie64a,JanZie66a}. 
Then Bialynicki-Birula and Mycielski suggested an additive nonlinear extension 
of the Schr\"odinger equation \cite{BiaMyc75a,BiaMyc76a}, that was treated in 
detail with more traditional concepts by Weinberg \cite{Wei89a1}. Later 
Bialynicki-Birula researched the Weyl equation and also the quantum mechanics 
of massless particles with the help of the hydrodynamic form 
\cite{Bia95a,Bia96a,Bia96c}. Kuz'menkov and Maksimov researched fermion systems 
providing a statistical background for the hydrodynamic view 
\cite{KuzMak99a,AndKuz07a}. 
The connection of vortices in quantum fluids and electromagnetism has been 
explored by Bialynicki-Birula and Bialynicka-Birula  
\cite{BiaBia71a,BiaEta00a,BiaBia03a} and recently related predictions were 
confirmed by experiments in slow ion-atom collisions 
\cite{OvcEta14a,SchmiEta14a}. The first relativistic extension is due to 
Takabayashi \cite{Tak57a}. Jackiw and coworkers proved that quantum field 
theories can be reformulated in a hydrodynamic form \cite{BisEta03a,JacEta04a}.

With this enumeration we want to express that in our opinion different 
interpretations may be mind provoking and hence useful \cite{BenEta14a}. This 
is also valid for the 
Bohm potential based approach \cite{Boh52a,Boh52a1,Hol93b,Wya05b}, that we 
consider closely related to the hydrodynamic one \cite{VanFul06a}. It leads to 
important observations, too, e.g. recently in the understanding of quantum 
tunneling \cite{HeiEta13a}. 

Here we would like to explore an important aspect of the use of the 
magnitude and phase of the complex wave function field, $\varphi$. Being 
interested in a splitting of the fundamental quantum {mechanical} equation into 
a ''classical'' and a ''quantum'' part namely, the representation 
\be
 \varphi = R \, e^{\frac{i}{\hbar}\alpha}
\ee{BASICREP}
is of genuine use. Here $\alpha$ plays the role of the classical action
for the corresponding classical dynamics and the canonical classical
momentum and energy are derived accordingly as
\be
  E = -\pd{\alpha}{t}, \qquad P = \nabla \alpha.
\ee{CLASS_ACTION}
The Schr\"odinger equation in its well-known form,
  \be
  -\frac{\hbar^2}{2m} \nabla^2 \varphi + V(x) \varphi = i\hbar \pd{}{t} \varphi,
  \ee{SCHRODINGER}
then can be rewritten in terms of the classical momentum, energy and the
quantum factor $R$ by observing the following derivatives:
 \be
 \pd{}{t}\varphi = \left(\frac{1}{R}\pd{R}{t}-\frac{i}{\hbar} \, E \right)\varphi, 
   \qquad 
   \nabla \varphi = \left(\frac{1}{R}\nabla R +\frac{i}{\hbar} \, P \right)\varphi.
 \ee{DERIVATIVES}
The Laplacian becomes
 \be
 \nabla^2\varphi = \left[ \nabla\left(\frac{\nabla R}{R}+\frac{i}{\hbar}P \right)+\left(\frac{\nabla R}{R}+\frac{i}{\hbar}P\right)^2 \right]\varphi.
 \ee{LAPLACIAN_PHI}
Now the Schr\"odinger equation (\ref{SCHRODINGER}) 
is separated into its real and imaginary parts as follows:
The real part connects the classical energy and momentum according to the 
classical formula, $E=P^2/2m$, and reveals a quantum correction, called the 
Bohm potential \cite{Boh52a,Boh52a1,Boh53a,Hol93b}:
 \be
 E = V -\frac{\hbar^2}{2m} \left[\nabla\frac{\nabla R}{R} + \left(\frac{\nabla R}{R}\right)^2 - \frac{P^2}{\hbar^2}\right]
 \ee{REAL_SCHROD}
The interpretation of this energy expression (\ref{REAL_SCHROD}) as a sum of a 
classical energy and a quantum modification,
 \be
 E = \left( \frac{P^2}{2m} + V \right)  \: - \frac{\hbar^2}{2m} \frac{\nabla^2 R}{R}, 
 \ee{ENERGY}
reveals a position-dependent quantum correction to the classical energy, $E$.
The imaginary part on the other hand leads to a first order time-evolution constraint
equation, 
 \be
 \frac{i\hbar}{R}\pd{R}{t} =  -\frac{\hbar^2}{2m} \frac{i}{\hbar}\left[\nabla P + \frac{2}{R} P\cdot\nabla R\right].
 \ee{IMAG_SCHROD}
Simplifying this imaginary part leads to an analogue of the mass density continuity equation,
 \be
   m\pd{R^2}{t} \, + \, \nabla\left(R^2 \, P \right)   = 0.
 \ee{CONTINUITY}
Upon introducing the velocity field via $P=mv$ and 
the local fluid density $\rho=R^2=|\varphi|^2$, this relation was interpreted as a continuity equation for the mass current
carried by a ''Madelung fluid''
\be
   \pd{\rho}{t} \, + \, \nabla\left(\rho \, v \right)   = 0.
\ee{MADELUNG_FLUID}
Similarly equation \re{ENERGY} is a ``Bernoulli equation'' of the corresponding 
rotation free momentum balance of a special Korteweg fluid 
\cite{VanFul06a,Van09a1} or an off mass shell relation \cite{FulKat98a}. From 
now on we follow the classical -- quantum separation hint by using the 
variables $R$ and $\alpha$. At the end we shall realize that exactly this 
splitting makes it possible to identify a conformal part in the quantum 
dynamics of the massive particles having a traceless contribution to the 
energy-momentum tensor.

\subsection{Schr\"odinger equation from action principle}

According to Schr\"odinger's original article about his equation
the following Action Principle can be formulated: instead of fulfilling the classical
Hamilton-Jacobi equation \cite{Sch23a}, it is violated so that its space-time integral
weighted by $|\varphi|^2$ achieves a variational extremum. The use of this
form of the weighting factor may be argued for by noting that only this leads to
a linear Euler-Lagrange variational equation.
The Quantum Action Principle behind the Schr\"odinger equation is given by
\cite{Sch26a,Sch26a1,Sch26a2}
   \be
   {\mathfrak S} = \int \left(\pd{S}{t} + \frac{|\nabla S|^2}{2m} +V \right) |\varphi|^2 \,  d^3x \: dt.
   \ee{SCHR_ACTION}
It has been interpreted  via a ''Boltzmannian'' eikonal ansatz: $S = 
\frac{\hbar}{i} \ln \varphi$. Using this ansatz leads to the following complex 
bilinear form of the quantum action:
\be
   {\mathfrak S} = \int \left[\frac{\hbar}{i}\varphi^* \pd{\varphi}{t} +\frac{\hbar^2}{2m}\nabla\varphi^* \cdot \nabla\varphi + V \varphi^* \varphi \right] d^3x \: dt
   \ee{ACTION_VARPHI}
Finally variation against $\varphi^*$ delivers the well-known Schr\"odinger equation,
linear in the complex wave function $\varphi$:
\be
    \pv{{\mathfrak S}}{\varphi^*} = \frac{\hbar}{i}\pd{\varphi}{t} - \frac{\hbar^2}{2m}\nabla^2 \varphi + V\varphi = 0
\ee{SCHRDEQ_AS_VARI}
It is straightforward to check by variation against $\varphi$
followed by a complex conjugation that the eikonal coefficient, $\hbar/i$, has to be pure imaginary.

Now we re-investigate this quantum action with magnitude-phase variables in order to
see the effect of the quantum -- classical splitting considered in the previous subsection.
Indeed the action also splits into quantum and classical parts,
\be
{\mathfrak S} = \int \left[ \frac{\hbar^2}{2m} \left(\nabla R\right)^2 + 
R^2 \left(\pd{\alpha}{t} + V + \frac{(\nabla\alpha)^2}{2m} \right) \right] d^3x \: dt
\ee{SCH_ACT_MAD}
The characteristic Lagrangian structure contained in this Quantum Action Principle can be summarized as follows:
$${\cal L} =  \hbar^2 \: ({\rm quantum \quad kinetic}) \,\, + \, \, 
	R^2 \: ({\rm classical \quad Hamilton-Jacobi \quad equation})$$
Finally we make some remarks about the relation between the pure classical action, $\alpha$ and
the complex action variable in the eikonal form, $S$ (more commonly used in derivations).
In fact one realizes that $ S = \alpha - i\hbar\ln R $, i.e. the real part of $S$ is the
classical $\alpha$. Certainly for $R=1$ the classical dynamics is recovered.
In the quantum propagation of massive objects, however, $\alpha$ and its derivatives,
$E$ and $P$, are not constants, their evolution couples to that of $R(x,t)$ exactly
via the Schr\"odinger equation. This fact typically reflects deviations from the
classical momentum and energy, and - as we shall see - also from the
on-mass-shell dispersion relation.

\section{Special Relativistic Quantum Mechanics}

{In this section we present the quantum -- classical splitting in the
above terms for the special relativistic free Klein-Gordon theory. Although, in fact,
the probability density interpretation is no more available in this case, a conserved
current and the corresponding continuity equation is easily derived. Restricing to
a single particle with mass $m$, this continuity reflects a content similar to the
one in the previous section. The corresponding four-velocity is, however, either not
normalized to one, and therefore is not a physical velocity, or its quantum,
''off-mass-shell'' contribution has to be splitted away from the classical part.}

\subsection{Klein-Gordon Lagrangian}

Disregarding the spin of the electron, the Schr\"odinger equation can be viewed as the
non-relativistic approximation to the Klein-Gordon equation -- in analogy to the
non-relativistic approximation to the relativistic Hamilton-Jacobi equation based on
the energy-momentum dispersion relation of the mass point $m$. Although the Klein-Gordon
equation does not describe the quantum energy levels of the H-atom precisely, for the study
of a quantum -- classical splitting it is more suitable due to its simplicity.
The quantum action is based on the Lagrange density
\be
{\cal L} =-\frac{1}{2} \partial_\mu \psi^* \, \partial^\mu \psi \, -
\frac{1}{2} \left(\frac{mc}{\hbar} \right)^2 \psi^* \, \psi,
\ee{KG_LAG}
containing a complex $\psi(x,t)$ field.
The action is a Lorentz-invariant integral,
\be
{\mathfrak S} = \int {\cal L} \: d^4x,
\ee{KG_ACTION}
with $dx^4=(cdt,d\vec{r})$.
We use physical units in which  $[{\cal L}] = $ energy density/c = $[ mc/L^3 ]$ and the Lorentz form $diag(1,-1,-1,-1)$.
In order to keep the relation to the non-relativistic wave function description on the one hand and to the classical relativistic mass point action (Maupertuis action) on the other hand, we include some further factors. The complex scalar field related to the wave function is written as 
\be
 \psi = \frac{\hbar}{\sqrt{mc}} \: R \, e^{\frac{i}{\hbar}\alpha}.
\ee{PSI_DEF}
Here $\alpha$ is the (real) classical action, as used in the previous section.  
The physical units of $R$ can be obtained from the mass term in the Lagrange density:
\be
\left(\frac{mc}{\hbar}\right)^2 \psi^* \psi = mc R^2
\ee{R_NORM}
is part of ${\cal L}$,  so it follows that $R^2$ is a {number density}.
Comparing this with the Maupertuis action for a classical mass point:
\be
 -\frac{1}{2} \int \left(\int mc^2 R^2 d^3x \right) dt = - \int mc^2 d\tau
\ee{MAUP}
{one would consider}
$\int R^2 \, d^3x \, = \, 2.$
{This step is of course not compulsory, one may re-interpret the scalar 
Klein-Gordon field as in the quantum field theory, representing an undetermined 
number of particles or more generally an undetermined value of mass, $M$. By 
doing so one reinterprets the Maupertuis action in eq.\re{MAUP} as -- 
$\int{Mc^2d\tau}$ rendering $M=m$ to be a very particular choice. In this case 
$\int{R^2d^3x = 2 M/m}$.  In the quantum interpretation here nothing excludes 
negative $M$-s, giving rise to Dirac's problem on describing holes in negative 
energy continua. For our purpose this debate is irrelevant; since we are only 
seeking  a motivation for the optimal quantum -- classical splitting for a 
given positive mass object.}

We consider now the derivatives of the complex field, $\psi$ in Lorentz-covariant notation.
The first derivative of $\psi$ is given by
\be
\partial_\mu \psi = \left(\frac{\partial_\mu R}{R} + \frac{i}{\hbar}\partial_\mu \alpha \right) \, \psi.
\ee{PSI_DERIV}
The derivative of the classical action is again a classical four-momentum and a four-velocity field
also can be introduced analogous to the non-relativistic treatment:
\be
 P_\mu = \partial_\mu \alpha, \qquad u_\mu = P_\mu/(mc).
\ee{MOM_DEF}

\subsection{Action principle with Madelung variables}

The special relativitic quantum action of the free massive particle can again be splitted
into a classical and a quantum part by using the magnitude-phase variables.
As a functional of the fields $R(x)$ and $\alpha(x)$ it reads as
\be
{\mathfrak S} = \frac{\hbar^2}{2mc} \int \left[\partial_\mu R \, \partial^\mu R \, + 
    \, \frac{R^2}{\hbar^2}\left(\partial_\mu\alpha \, \partial^\mu\alpha - (mc)^2 \right) \right] \, d^4x.
\ee{KG_MAD_ACTION}
Rewriting this expression, it is transformed into
$\hbar^2$ times quantum kinetic plus $R^2$ times classical part:
\be
{\mathfrak S} = \int \left[ 
  {\frac{\hbar^2}{2mc} \partial_\mu R \, \partial^\mu R} 
   \, \, + \, \, 
  { \frac{R^2}{2mc} \left( P_\mu \, P^\mu - (mc)^2 \right) } 
  			\right] \, d^4x 
\ee{Q_CLASS_MAD}
Now the classical part is the relativistic energy-momentum mass-shell expression, which
is classically zero, but in the quantum mechanics in general it differs from zero,
unless $R$ is a constant.
As it is well-known the Klein-Gordon action possesses a $U(1)$ phase symmetry of the $\psi(x)$
field. The corresponding U(1) Noether current is given by 
\be
   -J^\mu = \frac{i}{2\hbar} \left(\psi \: \partial^\mu\psi^* - \psi^* \partial^\mu\psi \right) =
   \frac{1}{mc} R^2 P^\mu = R^2 u^\mu
 \ee{DEF_U1}
constituting a number density 4-current $R^2u^\mu=\rho u^\mu$ based on the fluid picture.
It is interesting to realize that the variation of 
the quantum action ${\mathfrak S}$ with respect to the classical action (phase) $\alpha$ results in the conservation of this current:
\be
\frac{\delta {\mathfrak S}}{\delta \alpha} = - \partial_\mu \left(\frac{1}{mc} R^2 \partial^\mu \alpha \right) 
 = \partial_\mu J^\mu = 0.
\ee{VAR_ALFA}

For our present seek for the quantum - classical splitting of the content of quantum physics it is,  however, more important to study the other Euler-Lagrange equation of motion, the one obtained by variation against $R$. It delivers
 \be
 \frac{\delta {\mathfrak S}}{\delta R} = 
 		-\frac{\hbar^2}{mc} \Box R + \frac{R}{mc} \left(P_\mu P^\mu-(mc)^2 \right)=0.
 \ee{RVAR_KG}
Here $\partial_\mu^\mu = \Box$. This equation constitutes an off-mass shell dispersion relation for the classical 4-momentum
 \be
   P_\mu P^\mu -(mc)^2 = \hbar^2 \frac{\Box R}{R} . 
 \ee{OFF_SHELL}
Either one interprets this as quantum effects causing the free scalar field be off-mass shell even without any further interaction, or one speculates that perhaps the underlying
space-time metric recieves corrections if $R(x)$ is not a constant. In the latter case
we consider a metric view:
 \be
 g_{\mu\nu}u^\mu u^\nu = 1 + \left(\frac{\hbar}{mc}\right)^2 \frac{\Box R}{R} . 
 \ee{METRIC_VIEW}
This is a Compton wavelength scaled, locally Lorentzian spacetime metric.
{Although this observation does not enforces a quantum origin of the space-time 
metric itself, and therefore our suggestion for the classical -- quantum 
splitting in general has no intersection with quantum gravity theories, this 
metric view calls the attention to the fact that a certain handling of the 
quantum nature of the source term alone may modify the Einstein equation. This 
will be the basis
of our starting point in section \ref{GENREL}.}

\subsection{Klein-Gordon Energy-Momentum tensor}

In order to execute {the above outlined} 
program, one has to investigate the source term of gravity,
the energy-momentum tensor, more closely. So, before turning to the Einstein equation, we turn to the calculation of the Klein-Gordon energy-momentum tensor. First we review the textbook derivation \cite{ItzZub80b}, the one using $\psi$ and $\psi^*$.
{We note already here that in the context of general relativity the
energy-momentum tensor is obtained from the variation of the action against the
metric tensor, not like below. However, the result of our final choice on fixing the freedom of adding a total divergence to the Lagrange density and likewise a divergenceless
contribution to the energy-momentum tensor, will be in accord to the classical definition.}
{Without considering general relativity,} 
as a first step, the canonically conjugated complex ''momentum'' field is obtained,
 \be
 \Pi_\mu = \pv{{\cal L}}{\partial^\mu \psi} = \frac{1}{2} \partial_\mu \psi^*,
 \ee{CONJUG_MOM}
and then according to the familiar Legendre-transformation-like definition
the following energy-momentum tensor is presented:
 \be
 T_{\mu\nu} = \Pi_\mu \partial_\nu\psi + \Pi_\mu^* \partial_\nu \psi^* - g_{\mu\nu}{\cal L}.
 \ee{TIJ}
This can be rewritten in terms of $R$ and $\alpha$ as follows:
 \ba
 T_{\mu\nu} = mc R^2 w_{\mu\nu} + \frac{\hbar^2}{mc} U_{\mu\nu},
 \nl
 w_{\mu\nu} = u_\mu u_\nu-\frac{1}{2}g_{\mu\nu}(u_\alpha u^\alpha-1),
 \nl
 U_{\mu\nu} = \partial_\mu R \, \partial_\nu R - \frac{1}{2}g_{\mu\nu}  \partial_\alpha R \, \partial^\alpha R .
 \ea{RalfaTIJ}
 Here we note that the term proportional to $(u_\alpha u^\alpha-1)$ is also of quantum nature, in the order
 of $\hbar^2$. The only classical contribution to $T_{\mu\nu}$ is therefore $mc R^2 u_\mu u_\nu$, that of the
 dust consisting of point-particles with mass $m$ moving on Bohm trajectories according to
 the velocity field $u_\mu(x)$.
Replacing back the off-mass-shell relation (\ref{METRIC_VIEW}) into this expression leads to:
\be
 T_{\mu\nu} = mc R^2 u_\mu u_\nu  + 
 \frac{\hbar^2}{2mc} \left(2\partial_\mu R \, \partial_\nu R -g_{\mu\nu} 
 \left(\partial_\alpha R \partial^\alpha R + R\Box R \right) \right).
\ee{OFFSHELL_TIJ}
Here the ${\cal O}(\hbar^2)$ part is the quantum contribution, the rest is classical {dust}.
There are, however, other derivations of the energy-momentum tensor with a formally 
different result \cite{Del04m,Del13m,Car07m}. In the next subsection we explore the 
differences.

\subsection{Generalized Bohm-Takabayashi Energy-Momentum Tensor}

Although the Bohm-Takabayashi energy-momentum tensor \cite{Tak52a,Tak53a} was originally derived in the Madelung fluid picture, its validity is independent of the fluid interpretation. To begin with one takes the derivative of the off-mass-shell eqation (\ref{METRIC_VIEW}) and multiplies it by $R^2/2$:
 \be
 \frac{R^2}{2} \, \partial_\mu  \left[ u_\nu u^\nu -1-\frac{\hbar^2}{(mc)^2} \frac{\Box R}{R}  \right]  = 0.
 \ee{DERIV_METRIC_VIEW}
Introducing now the Compton wavelength $L_C=\hbar/mc$ and expanding the derivative of $u_\nu u^\nu$ we obtain
 \be
 R^2u^\nu\partial_\mu u_\nu - \frac{1}{2}L_C^2R^2 \partial_\mu\left(\frac{\Box R}{R}\right) = 0.
 \ee{EXPAND_DERIV}
One utilizes also the following identity (the Madelung fluid is irrotational)
  \be
  \partial_\mu u_\nu = \frac{1}{mc} \partial_\mu \partial_\nu \alpha =
  \frac{1}{mc} \partial_\nu \partial_\mu \alpha = \partial_\nu u_\mu.
  \ee{IRROT}
Therefore
  \be
  R^2u^\nu\partial_\mu u_\nu = R^2u^\nu\partial_\nu u_\mu = \partial_\nu (R^2u^\nu u_\mu) - u_\mu\partial_\nu(R^2u^\nu)
  \ee{DUE_TO_CONT}
  and due to continuity (eq.\ref{CONTINUITY}) the last term vanishes.
By these manipulations we obtain
 \be
 \partial_\nu \left(R^2u^\nu u_\mu\right) = \frac{1}{2}L_C^2R^2 \partial_\mu\left(\frac{\Box R}{R} \right) .
 \ee{CONTI_USED}
Further use of the Leibniz rule in this formula leads to
 \be
   R^2\partial_\mu\left(\frac{\Box R}{R} \right) = R \Box \partial_\mu R  - \partial_\mu R \, \Box R
 = \partial^\nu \left( R\partial_\nu\partial_\mu R - \partial_\nu R \, \partial_\mu R\right) .
 \ee{LEIBNIZ_NOCHMAL}
This form already reveals a vanishing divergence of the {\em Bohm-Takabayashi} tensor
  \be
     {\cal T}_{\mu\nu} = mc R^2 u_\mu u_\nu - \frac{\hbar^2}{2mc}
     \left( R\partial_\mu \partial_\nu R - \partial_\mu R \, \partial_\nu R \right).
  \ee{TAKABAYASHI}
Obviously this expression differs from the Klein-Gordon one (\ref{OFFSHELL_TIJ}) by
  \be
  \Delta_{\mu\nu} = T_{\mu\nu} - {\cal T}_{\mu\nu} = 
  \frac{\hbar^2}{2mc} \left(\partial_\mu R \partial_\nu R +R\partial_\mu\partial_\nu R - 
  g_{\mu\nu} (\partial_\alpha R\partial^\alpha R + R\Box R)\right) .
  \ee{Tmunu_DIFFERENCE}
This difference does not spoil the energy-momentum conservation,
because it has a vanishing divergence.
We note that
  \be
  \left(g_{\mu\nu}\Box -\partial_\mu\partial_\nu \right) \frac{R^2}{2} =
  g_{\mu\nu}(\partial_\alpha R \partial^\alpha R + R\Box R) -\partial_\mu R \partial_\nu R - 
  R\partial_\mu\partial_\nu R.
  \ee{THIS_NOTE}
Using this identity one realizes that the difference between the familiar Klein-Gordon
and the Bohm-Takabayashi tensor,
  \be
  \Delta_{\mu\nu} = \frac{\hbar^2}{4mc} \left(\partial_\mu\partial_\nu-g_{\mu\nu}\Box \right) R^2,
  \ee{TRANSVERSE_DIFF}
has a vanishing divergence \cite{CalEta70a}.
This difference can also be written as a divergence of a three-index tensor,
$\Delta_{\mu\nu}=\partial^\alpha f_{\alpha\mu\nu}$
with
\be
f_{\alpha\mu\nu}=\frac{\hbar^2R}{2mc}\left(g_{\alpha\mu}\partial_\nu R-g_{\mu\nu}\partial_\alpha R \right).
\ee{THREEINDEX}
We note that in general it is allowed to add a term to the energy-momentum tensor
with vanishing divergence, such a term does not change the conservation.
Energy, however, has a physical meaning. In fact, the continous symmetries of the underlying
action govern the correct expression. The proper energy-momentum tensor can be obtained
by taking into account all continuous symmetries via their infinitesimal generators,
according to a procedure described in  \cite{ForRom04a,Pon11a}.
The difference $\Delta_{\mu\nu}$ is related to the realization of the conformal symmetry.
The full  energy-momentum tensor is the proper mixture of the above expressions. The 
general tensor contains a parameter $\lambda$ multiplying $\Delta_{\mu\nu}$ and added 
to the Bohm-Takabayashi tensor \re{TAKABAYASHI}:
\be
{\mathfrak T}_{\mu\nu}=mc R^2u_\mu u_\nu + {\mathfrak U}_{\mu\nu} + \lambda \Delta_{\mu\nu}.
\ee{GENERAL_TIJ}
The conformal part can be identified by inspecting the trace of the energy-momentum tensor,
\be
{\mathfrak T}_\mu^\mu = \left\{ 1 + L_C^2 \: \frac{1-3\lambda}{4} \Box \right\} 
(mc R^2). 
\ee{TRACE_OF_TMUNU}
For $\lambda=0$ one arrives at the original Bohm-Takabayashi tensor.
For $\lambda=1$ the original Klein-Gordon case emerges. Finally,
for $\lambda=1/3$ only classical dust contributes to the trace and all terms proportional to
$L_C^2\propto\hbar^2$ -- the quantum part of the energy-momentum tensor --
are altogether traceless. This will be the basis of the classical - quantum splitting
of the Einstein equation when considering classical gravity with quantum sources.
In order to prepare this study in the next section, we
express the generalized Bohm-Takabayashi energy-momentum tensor in scaling 
variables.
Using  $R=e^{\sigma}/\sqrt{V}$, where $V$ is constant, one easily gets
 \be
 {\mathfrak T}_{\mu\nu} = \frac{mc}{V} e^{2\sigma} u_\mu u_\nu + 
 \frac{\hbar^2}{2mcV} \, e^{2\sigma} {\mathfrak W}_{\mu\nu},
 \ee{TIJ_SCALING_VARS}
with
 \be
 {\mathfrak W}_{\mu\nu} =  2\lambda \partial_\mu\sigma \, \partial_\nu\sigma +
 	(\lambda-1)\partial_\mu\partial_\nu\sigma - 
	\lambda\eta_{\mu\nu}\left(2\partial_\alpha\sigma\partial^\alpha\sigma + 
	\Box\sigma \right).
 \ee{SCALING_QUANTUM_PART}
 This expression readily reminds us to a dilaton field $\sigma(x)$ 
 \cite{BraDic61a,Fuj71a,Fuj72a,FujMae04b,GasVen03a,Bra05m}.

 \section{\label{GENREL} General Relativistic Quantum Mechanics}

In this section we turn to the theory of general relativity. In particular we 
suggest to utilize the generalized energy-momentum tensor  source term in the 
Einstein equation by using the entire ${\mathfrak  T}_{\mu\nu}$ given in eq. 
\re{TIJ_SCALING_VARS}. This supports the classical  source term with quantum 
contributions in general space-time:
\be
G^{curved}_{\mu\nu} = \frac{8\pi G}{c^3}
{T}^{curved}_{\mu\nu}.
\ee{EINSTEIN} 

With the help of a conformal transformation we shall obtain an equivalent form of this 
equation as 
{\be
G^{flat}_{\mu\nu} - \Lambda \eta_{\mu\nu} =  \frac{8\pi G}{c^3} 
T^{remnant}_{\mu\nu}.
\ee{FLATEinstein}
}

Here $T^{remnant}_{\mu\nu}$ shall have an essentially reduced form 
relative to the generalized Bohm-Takabayashi energy-momentum tensor, cf. eq. 
\re{TIJ_SCALING_VARS}, discussed in the previous section. We design a conformal 
factor, $e^{2s}$, to compensate the $R^2 = e^{2\sigma}/V$ factor 
discussed previously. By doing so further effects arise, among others a cosmological
constant like source term. Finally we estimate, that if gravity had caused a 
quantum binding, in what mass range the pairwise bound objects should fall in 
order to count quantitatively for the cosmological constant effect observed to 
day.

\subsection{Conformal transformation of the Einstein equation}

Our starting point is the Einstein equation eq. \re{EINSTEIN}  with the curved 
space-time version of the generalized Bohm-Takabayashi energy-momentum tensor 
as a source term. We introduce a conformal transformation that flattens out the 
metric of the Einstein equation, considering the curved geometry with a 
conformal metric tensor 
\be
g^{curved}_{\mu\nu} = e^{2s}\eta_{\mu\nu},
\ee{confmetr}
with $\eta_{\mu\nu}$ being the Minkowski metric and $s(x)$ a scalar function of 
the space-time coordinates. A conformal transformation of the energy-momentum 
tensor is given by 
\footnote{It is so in order to transform a traceless tensor to another 
traceless one by a conformal transformation.}
\be
{T}^{curved}_{\mu\nu} = e^{-2s}{\mathfrak T}_{\mu\nu}
\ee{confenergymom}
and that of the Einstein tensor by \cite{Wal84b}
\be
G^{curved}_{\mu\nu} = G_{\mu\nu} + 
	2\partial^\mu s \, \partial_\nu s  -2\partial^\mu\partial_\nu s +
\eta_{\mu\nu}\left(2\Box s + \partial_\alpha s \partial^\alpha s \right).
\ee{CONFEINSTEIN}
Quantities and partial derivatives on the right hand side refer to the flat 
metric 
$\eta_{\mu\nu}$. 
Substituting  \re{TIJ_SCALING_VARS}, \re{SCALING_QUANTUM_PART}, \re{confenergymom} and 
\re{CONFEINSTEIN}  into eq. \re{EINSTEIN} we arrive at 
\begin{gather}
G_{\mu\nu} + 
	2\partial_\mu s\, \partial_\nu s -
	2\partial_\mu\partial_\nu s+
	\left(2\Box s +  \partial_\alpha s \partial^\alpha s  
	\right)\eta_{\mu\nu}  =\nonumber\\
   \frac{8\pi G}{c^3} e^{-2s} \frac{\hbar^2}{2 m c V}e^{2\sigma}\left[
2\lambda \partial_\mu\sigma \, \partial_\nu\sigma +
 	(\lambda-1)\partial_\mu\partial_\nu\sigma - 
	\lambda\eta_{\mu\nu}\left(2\partial_\alpha\sigma\partial^\alpha\sigma + 
	\Box\sigma \right)\right].
\label{cEin}\end{gather} 
It is obvious, that the $s=\sigma$ choice represents the optimal reduction 
formula. The very same choice unifies several terms included in the curved Einstein 
and the generalized Bohm-Takabayashi tensors.

\subsection{Identifying the quantum effects}

For the sake of deriving a most simple expression  we may chose the trace 
parameter, $\lambda$ and the wave function normalization volume $V$, so that 
the terms containing tensorial forms of partial derivatives compensate each 
other. This requires the equality of the coefficients of the terms 
$\partial_\mu s\partial_\nu s$ and $\partial_\mu\partial_\nu s$ on both sides 
of eq. \re{cEin}:
\begin{eqnarray}
2 &=& 2 \lambda \frac{8\pi G}{c^3}\frac{\hbar^2}{2mcV}, \label{c1}\\
-2 &=& (\lambda -1) \frac{8\pi G}{c^3}\frac{\hbar^2}{2mcV}. \label{c2}
\end{eqnarray} 
The solution of these requirements fixes $\lambda = 1/3$, the same value which 
leaves the classical part of the trace in the dust form ${\mathfrak T}^\mu_\mu 
= mc R^2$, cf. eq. \re{TRACE_OF_TMUNU}. The optimal normalization volume becomes
\be
 V = \frac{4\pi}{3} L_S L_C^2,
\ee{QUANT_VOLUME}
upon using the Schwarzschild length by $L_S = Gm/c^2$ and the Compton 
wavelength 
$L_C=\hbar/mc$. We note that $V$ is the Planck volume scaled by $M_p/m$.
The remaining terms in eq. \re{cEin} can be collected into an Einstein equation 
containing a cosmological term
\be
G_{\mu\nu} -\Lambda \eta_{\mu\nu} = \frac{8\pi G}{c^3} \frac{m}{V} u_\mu u_\nu.
\ee{flatEin}
We note that  $8\pi Gm/(c^3 V) = 8 \pi L_S/V = 6/L_C^2$.

Here we have identified a {\em cosmological term} proportional to the 
off-mass-shell effect on the Bohm trajectories:
\be
\Lambda = -3 \left(\Box \sigma + \partial_\mu\sigma\partial^\mu\sigma \right) = -3 \frac{\Box R}{R}.
\ee{COSMO}
This result delivers a key to a new thinking about the cosmological constant.
In this scenario quantum effects, in particular an attractive interaction, may 
lower the effective $P_\mu P^\mu$ for a particle below the classical mass shell 
value, acting this way as an effective cosmological constant (cf. 
eq.\ref{OFF_SHELL}):
\be
\Lambda = \frac{3}{\hbar^2} \, \left( (mc)^2 - P_\mu P^\mu \right).
\ee{LAMBDA_OFF_SHELL}

Solutions of the quantization volume \re{QUANT_VOLUME} and on the induced 
cosmological constant \re{COSMO} might indicate a possible agreement with the 
scale-factor duality idea \cite{GasVen03a}.

\subsection{{Cosmological effect from quantum binding in gravitational 
$-\alpha/r$ potential}}

Based on this view we would like to make a new order of magnitude estimate for 
the source of a cosmological constant. A plane-wave solution of the 
Klein-Gordon equation leads to a zero cosmological term.  Therefore we 
introduce a potential $A^\mu$ in the Klein-Gordon equation and look for bounded 
solutions. Then it is easy to see, that the modification of the previous train 
of thought requires the modification of the momentum $P^\mu = \partial^\mu 
\alpha -A^\mu$,  the Einstein-equation \re{EINSTEIN} does not change, therefore 
the cosmological term is the same. Then assuming a particular reference frame 
and introducing a Coulomb like potential $A^\mu=(V(r),0^i)$, with $V(r) = 
-\alpha/r$ we obtain the following estimation for the cosmological term
  \be
   \Lambda = 3 \, {\frac{\nabla^2 R}{R}} = 3 \, \left( \frac{1}{a^2} - \frac{2}{ar} \right)
  \ee{BOHR_RADIUS}
with $a=L_C/\alpha$ being the {Bohr radius} \cite{GalPas90b1,Don11b}. Here the constant part must belong to the cosmological effect, while the $-1/r$ like part to the potential energy term in the general non-relativistic Schr\"odinger equation. The relativistic treatment does not change the $r \rightarrow \infty$ limit.

The quantum energy part is a spatial constant, $3/a^2$ which may be in the correct order of magnitude.
 
In $c=1$ units the gravitational Newtonian potential has the coupling constant
$\alpha=Gm^2/\hbar=(m/M_P)^2$ between two mass $m$ objects. The corresponding 
Bohr radius 
amounts to $a_B=\hbar/m\alpha=L_P(M_P/m)^3$. Because in an equal mass ''gravonium'' 
the reduced mass is $m/2$, the radius we count with is $a=2a_B$. This leads to the
following estimate
\be
\Lambda = \frac{3}{a^2} = \frac{3}{4L_P^2} \left(\frac{m}{M_P} \right)^6.
\ee{GRAVONIUM}
Using known values for $M_P$, $L_P$ and $L_P^2\Lambda = 2.56 \cdot 10^{-122}$ 
one arrives at $m \approx 68$ MeV. This accounts to a total gravonium mass of $2m \approx 138$ MeV.

\section*{Summary}

In summary we have explored the classical -- quantum splitting of the 
Schr\"odinger equation by using the magnitude-phase representation of the 
complex wave function. By doing so not the Madelung fluid interpretation, but 
the partial conformal symmetry hidden in the relativistic Klein-Gordon 
Lagrangian, a simple relativistic generalization behind the Schr\"odinger 
Quantum Action, was in focus. Although the mass term breaks conformal 
invariance, in the limit of zero mass the rest of the theory should restore 
this. Accordingly the determination of the proper energy-momentum tensor has to 
take this symmetry into account.

Following the general mathematical recipe \cite{ForRom04a,CalEta70a,GotEta92a}, 
we concluded that neither the naive expression - frequently found in textbooks 
- nor the Bohm-Takabayashi form of $T_{\mu\nu}$ takes care of this symmetry. A 
conformal transformation of the Einstein tensor can be carried out which 
separates a classical fluid-like contribution of the free particle field to the 
classical gravity from quantum corrections in the energy-momentum tensor, 
$T_{\mu\nu}$, by assuming a simple (in the $\hbar=0$ limit vanishing) 
modification of the Einstein equation.

Moreover this quantum -- classical splitting of the source term of the Einstein 
equation functions only if the Bohm potential part of $T_{\mu\nu}$ is traceless 
($\lambda=1/3$). Beyond this a cosmological term arises which was found to be 
proportional to the off-mass-shell measure of particles moving on Bohmian 
trajectories ($\Lambda=-3\Box R / R$). As a small bonus the natural reference 
quantization volume belonging to the normalization of the total mass $M$ 
represented by the scalar field, ($R=|\varphi|$, $\int R^2 d^3x = 2M/m$), 
becomes a Planck-scale based quantity ($V= \frac{4\pi}{3}L_SL_C^2$).  In fact 
we constructed a particular Jordan-Einstein  frame change 
\cite{Dic62a,Fla04a,FarNad07a}, which is optimally suited to simplify leading 
order quantum source effects.

Finally we investigated the  induced cosmological term in case of  quantum 
bound states in a simple, static $-\alpha/r$ Newtonian gravitational potential 
of two mass $m$ scalar objects. Due to eq.(\ref{GRAVONIUM}) the estimate for 
the total gravonium mass is $138$ MeV. Since such objects - if they exist - are 
only superweakly bound (by gravity only), they cannot be mixed with ordinary 
matter.

As our first conclusive remark we note that recent approaches of quantum 
geometry recognize the connection to  conformal transformation and Weyl 
geometry from various points of view. For example Carroll \cite{Car10b,Car10m} 
reviews many different works in this respect. On the other hand Koch summarizes 
and further elaborates some issues regarding reservations of nonstandard 
quantum interpretations \cite{Koc10p,Koc09m1} (see also \cite{Smo06m, Schme11m} 
related specifically to the mentioned work of Wallstrom \cite{Wal94a}). Our 
treatment is based on energetic considerations and focuses on the clear, 
formal, (universal) mathematical aspects, trying to avoid the traps of 
interpretational issues. 

Our second remark corresponds to the trace anomaly of quantum field theories. 
At first sight our Klein-Gordon quantum mechanics has nothing to do with an 
effect that emerges in quantum fields in curved space-time \cite{CapLau11a}. 
The optimal choice $\lambda = 1/3$, removes the trace of the quantum part and 
leaves the classical part.  Trace anomaly on the other hand usually occurs if 
quantum effects lead to nonvanishing  $T^\mu_\mu$ corrections to a classically 
traceless energy-momentum tensor. Our approach presented in this paper does not 
suit to the classification scheme of Flanagan \cite{Fla04a}: our scalar field 
variable $\sigma$ emerges as the quantum (non Hamilton-Jacobi) part of the 
action.

The essential difference lies in the separation of quantum parts. For the 
Schr\"odinger equation the Madelung variables \re{BASICREP} are the classical 
real action $\alpha$ and the $\sigma = \ln R + \ln V /2$  characterizing the 
probability amplitude, 
$$
\psi = R \, e^{\frac{i}{\hbar}\alpha} = 
	 \frac{1}{\sqrt{V}}e^{\frac{i}{\hbar}\alpha + \sigma}= 
	\frac{1}{\sqrt{V}} e^{\frac{i}{\hbar}(\alpha - i \hbar \sigma)}.
$$
In a Feynman path integral formulation the transition probability is written as 
$$
\psi =  \psi_0  e^{\frac{i}{\hbar}S} =\frac{1}{\sqrt{V}}  
e^{\frac{i}{\hbar}(Re(S) + i Im(S))}
$$
where $S$ is the action. Comparing the two expressions one realizes that a 
loop-expansion in the Feynman formalism requires a resummation in the Madelung 
variables and vice versa. The comparison could be more rewarding with the 
hydrodynamic version of quantum field theories \cite{BisEta03a,JacEta04a}.

\section*{Acknowledgement}

We thank Manfried Faber for detailed discussions. Antal Jakov\'ac, Andr\'as 
Patk\'os and Reinhard Alkofer contributed with inspiring remarks at the ACHT 
(Austrian-Croatian-Hungarian Triangle) Meeting in Retzhof, June 2013. We also 
thank to the referees for the constructive remarks. This work was supported by 
the Hungarian National Research Fund OTKA K104260. 

\bibliographystyle{unsrt}

\begin{thebibliography}{10}

\bibitem{Sit16a}
W.~de~Sitter.
\newblock On {E}instein's theory of gravitation and its astronomical
  consequences.
\newblock {\em Monthly Notices of the Royal Astronomical Society}, 76:699--728,
  1916.

\bibitem{Sit16a1}
W.~de~Sitter.
\newblock Erratum: On {E}instein's theory of gravitation and its astronomical
  consequences. {S}econd paper.
\newblock {\em Monthly Notices of the Royal Astronomical Society}, 77:155--184,
  1916.

\bibitem{Sit17a1}
W.~de~Sitter.
\newblock Einstein's theory of gravitation and its astronomical consequences.
  {T}hird paper.
\newblock {\em Monthly Notices of the Royal Astronomical Society}, 78:3--28,
  1917.

\bibitem{Fri22a}
A.~Friedmann.
\newblock {\"U}ber die {K}r\"ummung des {R}aumes.
\newblock {\em Zeitschrif f\"ur Physik}, 10:377--386, 1922.

\bibitem{Fri24a}
A.~Friedmann.
\newblock {\"U}ber die {M}{\"o}glichkeit einer {W}elt mit konstanter negativer
  {K}r{\"u}mmung des {R}aumes.
\newblock {\em Zeitschrif f\"ur Physik}, 21(1):326--332, 1924.

\bibitem{Lem27a}
G.~E. Lema\^{\i}tre.
\newblock Un univers homog\'ene de masse constante et de rayon croissant
  rendant compte de la vitesse radiale des n\'ebuleuses extragalactiques.
\newblock {\em Annals of the Scientific Society of Brussels}, 47A:49--59, 1927.

\bibitem{Ein17a}
A.~Einstein.
\newblock Kosmologische {B}etrachtungen zur allgemeinen
  {R}elativit\"atstheorie.
\newblock {\em Sitzungsberichte der K\"oniglich Preu{\ss}ischen Akademie der
  Wissenschaften (Berlin)}, pages 142--152, 1917.

\bibitem{Ein31a}
A.~Einstein.
\newblock Zum kosmologischen {P}roblem der allgemeinen {R}elativit\"atstheorie.
\newblock {\em Sitzungsberichte der K\"oniglich Preu{\ss}ischen Akademie der
  Wissenschaften (Berlin)}, pages 235--237, 1931.

\bibitem{Pad03a}
T.~Padmanabhan.
\newblock Cosmological constant - the weight of the vacuum.
\newblock {\em Physics Report}, 380:235--320, 2003.
\newblock arXiv:hep-th/0212290.

\bibitem{Bur13m}
C.P. Burgess.
\newblock The cosmological constant problem: Why it' s hard to get dark energy
  from micro-physics.
\newblock 2013.
\newblock arXiv:1309.4133.

\bibitem{Hub29a}
E.~Hubble.
\newblock A relation between distance and radial velocity among extra-galactic
  nebulae.
\newblock {\em Proceedings of the National Academy of Sciences of the United
  States of America}, 15(3):168--173, 1929.

\bibitem{CarEta92a}
Press W.~H. Carroll, S.M. and E.L. Turner.
\newblock The cosmological constant.
\newblock {\em Annual review of astronomy and astrophysics}, 30:499--542, 1992.

\bibitem{Wei89a}
S.~Weinberg.
\newblock The cosmological constant problem.
\newblock {\em Reviews of Modern Physics}, 61(1):1--23, 1989.

\bibitem{Car01a}
S.M. Carroll.
\newblock The cosmological constant.
\newblock {\em Living Reviews in Relativity}, 3:1, 2001.

\bibitem{Planck13m}
et~al Ade, P.A.R.
\newblock Planck 2013 results. {XVI}. {C}osmological parameters.
\newblock 2013.

\bibitem{Pad06p}
T.~Padmanabhan.
\newblock Dark energy: Mystery of the millennium.
\newblock In J.-M. Alimi and F\H uzfa A., editors, {\em Albert Einstein Century
  International Conference}, AIP Conference Proceedings, pages 179--196.
  American Institute of Physics, 2006.
\newblock arXiv:astro-ph/0603114.

\bibitem{Sch26a}
E.~Schr\"odinger.
\newblock Quantisierung als {E}igenwertproblem ({E}rste {M}itteilung).
\newblock {\em Annalen der Physik}, 79:361--376, 1926.

\bibitem{Sch26a1}
E.~Schr\"odinger.
\newblock Quantisierung als {E}igenwertproblem ({Z}weite {M}itteilung).
\newblock {\em Annalen der Physik}, 79:489--527, 1926.

\bibitem{Sch26a2}
E.~Schr\"odinger.
\newblock {\"U}ber das {V}erh\"altnis der {H}eisenberg-{B}orn-{J}ordanschen
  {Q}uantenmechanik zu der meinem.
\newblock {\em Annalen der Physik}, 79:734--756, 1926.

\bibitem{Sch26a3}
E.~Schr\"odinger.
\newblock An undulatory theory of the mechanics of atoms and molecules.
\newblock {\em Physical Review}, 28(6):1049--1070, 1926.

\bibitem{Boh52a}
D.~Bohm.
\newblock A suggested interpretation of the quantum theory in terms of "hidden"
  variables {I}.
\newblock {\em Physical Review}, 85(2):166--179, 1952.

\bibitem{Boh52a1}
D.~Bohm.
\newblock A suggested interpretation of the quantum theory in terms of "hidden"
  variables {II}.
\newblock {\em Physical Review}, 85(2):180--193, 1952.

\bibitem{Boh53a}
D.~Bohm.
\newblock Proof that probability density approaches $|\phi |^2$ of the quantum
  theory.
\newblock {\em Physical Review}, 89(2):458--466, 1953.

\bibitem{BohVig54a}
D.~Bohm and J.~P. Vigier.
\newblock Model of the causal interpretation of quantum theory in terms of a
  fluid with irregular fluctuations.
\newblock {\em Physical Review}, 96(1):208--216, 1954.

\bibitem{Tak52a}
T.~Takabayasi.
\newblock On the formulation of quantum mechanics associated with classical
  pictures.
\newblock {\em Progress of Theoretical Physics}, 8(2):143--182, 1952.

\bibitem{Tak53a}
T.~Takabayasi.
\newblock Remarks on the formulation of quantum mechanics with classical
  pictures and on relations between linear scalar fields and hydrodynamical
  fields.
\newblock {\em Progress of Theoretical Physics}, 9(3):187--222, 1953.

\bibitem{Ein15a}
A.~Einstein.
\newblock Die {F}eldgleichungen der {G}ravitation.
\newblock {\em {S}itzungsberichte der K\"oniglich {P}reussische {A}kademie der
  {W}issenschaften zu Berlin}, pages 844--847, 1915.

\bibitem{Ein16a}
A.~Einstein.
\newblock Die {G}rundlagen der {A}lgemeine {R}elativit\"atstheorie.
\newblock {\em Annalen der Physik}, 4(49):769--822, 1916.

\bibitem{Hil15a}
D.~Hilbert.
\newblock Die {G}rundlagen der {P}hysik.
\newblock {\em Nachrichten von der Gesellschaft der Wissenschaften zu
  G\"ottingen, Mathematisch-Physikalische Klasse}, pages 395--408, 1915.

\bibitem{Mad26a}
E.~Madelung.
\newblock Quantentheorie in hydrodynamischer {F}orm.
\newblock {\em Zeitschrift f\"ur Physik}, 40:322--326, 1926.
\newblock in German.

\bibitem{Jan69a}
L.~J\'anossy.
\newblock The hydrodynamical model of wave mechanics, ({T}he many body
  problem).
\newblock {\em Acta Physica Hungarica}, 27:35--46, 1969.

\bibitem{Boe92a}
J.C.A. Boeyens.
\newblock The geometry of quantum events.
\newblock {\em Speculations in Science and Technology}, 15:192--210, 1992.

\bibitem{Wal94a}
T.~C. Wallstrom.
\newblock Inequivalence between the {S}chr\"odinger equation and the {M}adelung
  hydrodynamic equations.
\newblock {\em Physical Review A}, 49(3):1613--1617, 1994.

\bibitem{Jan62a}
L.~J{\'a}nossy.
\newblock Zum hydrodynamischen {M}odell der {Q}uantenmechanik.
\newblock {\em Zeitschrift f{\"u}r Physik}, 169(1):79--89, 1962.

\bibitem{JanZie63a}
L.~J\'anossy and M.~Ziegler.
\newblock The hydrodynamical model of wave mechanics {I}., ({T}he motion of a
  single particle in a potential field).
\newblock {\em Acta Physica Hungarica}, 16(1):37--47, 1963.

\bibitem{JanZie64a}
L.~J\'anossy and M.~Ziegler-N\'aray.
\newblock The hydrodynamical model of wave mechanics {II}., ({T}he motion of a
  single particle in an external electromagnetic field).
\newblock {\em Acta Physica Hungarica}, 16(4):345--353, 1964.

\bibitem{JanZie66a}
L.~J\'anossy and M.~Ziegler-N\'aray.
\newblock The hydrodynamical model of wave mechanics {III}., ({E}lectron spin).
\newblock {\em Acta Physica Hungarica}, 20:233--249, 1966.

\bibitem{BiaMyc75a}
I.~Bialynicki-Birula and J.~Mycielski.
\newblock Wave equations with logarithmic nonlinearities.
\newblock {\em Bull. Acad. Polon. Sci. Cl}, 3(23):461, 1975.

\bibitem{BiaMyc76a}
I.~Bialynicki-Birula and J.~Mycielski.
\newblock Nonlinear wave mechanics.
\newblock {\em Annals of Physics}, 100:62--93, 1976.

\bibitem{Wei89a1}
S.~Weinberg.
\newblock Testing quantum mechanics.
\newblock {\em Annals of Physics}, 194:336--386, 1989.

\bibitem{Bia95a}
I.~Bialynicki-Birula.
\newblock Hydrodynamic form of the weyl equation.
\newblock {\em Acta Physica Polonica}, 26(7):1201--1208, 1995.

\bibitem{Bia96a}
I.~Bialynicki-Birula.
\newblock Hydrodynamics of relativistic probability flows.
\newblock {\em Nonlinear Dynamics, Chaotic and Complex Systems}, pages 64--71,
  1996.

\bibitem{Bia96c}
I.~Bialynicki-Birula.
\newblock The photon wave function.
\newblock In {\em Coherence and Quantum Optics VII}, pages 313--322. 1996.

\bibitem{KuzMak99a}
L.~S. Kuz'menkov and S.G. Maksimov.
\newblock Quantum hydrodynamics of particle systems with {C}oulomb interaction
  and quantum {B}ohm potential.
\newblock {\em Theoretical and Mathematical Physics}, 118:227--240, 1999.

\bibitem{AndKuz07a}
P.A. Andreev and L.S. Kuz'menkov.
\newblock On equations for the evolution of collective phenomena in fermion
  systems.
\newblock {\em Russian Physics Journal}, 50(12):1251--1258, 2007.

\bibitem{BiaBia71a}
I.~Bialynicki-Birula and Z.~Bialynicka-Birula.
\newblock Magnetic monopoles in the hydrodynamic formulation of quantum
  mechanics.
\newblock {\em Physical Review D}, 3(10):2410--2412, 1971.

\bibitem{BiaEta00a}
I.~Bialynicki-Birula, Z.~Bialynicka-Birula, and C.~{\'S}liwa.
\newblock Motion of vortex lines in quantum mechanics.
\newblock {\em Physical Review A}, 61(3):032110, 2000.

\bibitem{BiaBia03a}
I.~Bialynicki-Birula and Z.~Bialynicka-Birula.
\newblock Vortex lines of the electromagnetic field.
\newblock {\em Physical Review A}, 67(6):062114, 2003.

\bibitem{OvcEta14a}
S.Y. Ovchinnikov, J.H. Macek, and D.R. Schultz.
\newblock Hydrodynamical interpretation of angular momentum and energy transfer
  in atomic processes.
\newblock {\em Physical Review A}, 90(6):062713, 2014.

\bibitem{SchmiEta14a}
L.~Ph.~H. Schmidt, C.~Goihl, D.~Metz, H.~Schmidt-B{\"o}cking, R.~D{\"o}rner,
  S.~Yu. Ovchinnikov, JH. Macek, and DR. Schultz.
\newblock Vortices associated with the wave function of a single electron
  emitted in slow ion-atom collisions.
\newblock {\em Physical Review Letters}, 112(8):083201, 2014.

\bibitem{Tak57a}
T.~Takabayasi.
\newblock Relativistic hydrodynamics of the {D}irac matter {P}art {I}.
  {G}eneral theory.
\newblock {\em Supplement of the Progress of Theoretical Physics}, (4):1--80,
  1957.

\bibitem{BisEta03a}
B.~Bistrovic, R.~Jackiw, H.~Li, V.~P. Nair, and S.-Y. Pi.
\newblock Non-{A}belian fluid dynamics in {L}agrangian formulation.
\newblock {\em Physical Review D}, (67):025013(11), 2003.
\newblock (hep-th/0210143).

\bibitem{JacEta04a}
R.~Jackiw, V.~P. Nair, S-Y. Pi, and A.~P. Polychronakos.
\newblock Perfect fluid theory and its extensions.
\newblock {\em Journal of Physics A}, 37:R327--R432, 2004.
\newblock arXiv:hep-ph/0407101.

\bibitem{BenEta14a}
A.~Benseny, G.~Albareda, A.~S. Sanz, J.~Mompart, and X.~Oriols.
\newblock Applied bohmian mechanics.
\newblock {\em The European Physical Journal D}, 68(10), 2014.

\bibitem{Hol93b}
P.~R. Holland.
\newblock {\em The Quantum Theory of Motion}.
\newblock Cambridge University Press, Cambridge, 1993.

\bibitem{Wya05b}
R.~E. Wyatt.
\newblock {\em Quantum dynamics with trajectories ({I}ntroduction to quantum
  hydrodynamics)}.
\newblock Springer, 2005.
\newblock Volume 28 in Interdisciplinary Applied Mathematics, contributions by
  Trahan, C. J.

\bibitem{VanFul06a}
P.~V\'an and T.~F\"ul\"op.
\newblock Weakly nonlocal fluid mechanics - the {S}chr\"odinger equation.
\newblock {\em Proceedings of the Royal Society, London A}, 462(2066):541--557,
  2006.
\newblock (quant-ph/0304062).

\bibitem{HeiEta13a}
D.M. Heim, W.P. Schleich, P.M. Alsing, J.~P. Dahl, and S.~Varr\'o.
\newblock Tunneling of an energy eigenstate through a parabolic barrier viewed
  from wigner phase space.
\newblock {\em Physics Letters A}, 377(31):1822--1825, 2013.

\bibitem{Van09a1}
P.~V\'an.
\newblock Weakly nonlocal non-equilibrium thermodynamics - variational
  principles and {S}econd {L}aw.
\newblock In Ewald Quak and Tarmo Soomere, editors, {\em Applied Wave
  Mathematics (Selected Topics in Solids, Fluids, and Mathematical Methods)},
  chapter III, pages 153--186. Springer-Verlag, Berlin-Heidelberg, 2009.
\newblock (arXiv:0902.3261).

\bibitem{FulKat98a}
T.~F\"ul\"op and S.~D. Katz.
\newblock A frame and gauge free formulation of quantum mechanics.
\newblock quant-ph/9806067, 1998.

\bibitem{Sch23a}
E.~Schr\"odinger.
\newblock {\"U}ber eine bemerkenswerte {E}igenschaft der {Q}uantenbahnen eines
  einzelnen {E}lektrons.
\newblock {\em Zeitschrift f\"ur Physik}, 12(1):13--23, 1923.

\bibitem{ItzZub80b}
C.~Itzykson and J.B. Zuber.
\newblock {\em Quantum field theory}.
\newblock McGraw-Hill, New York, etc., 1980.

\bibitem{Del04m}
D.H. Delphenic.
\newblock A geometric origin for the {M}adelung potential.
\newblock arXiv:gr-qc/0211065.

\bibitem{Del13m}
D.H. Delphenic.
\newblock A strain tensor that couples to the {M}adelung stress tensor.
\newblock arXiv:1303.3582.

\bibitem{Car07m}
R.~Carroll.
\newblock Remarks on geometry and the quantum potential.
\newblock arXiv:math-ph/0701007.

\bibitem{CalEta70a}
C.G. Callan, S.~Coleman, and R.~Jackiw.
\newblock A new improved energy-momentum tensor.
\newblock {\em Annals of Physics}, 59:42--73, 1970.

\bibitem{ForRom04a}
M.~Forger and H.~R\"omer.
\newblock Currents and the energy-momentum tensor in classical field theory.

\bibitem{Pon11a}
J.~M. Pons.
\newblock Noether symmetries, energy-momentum tensors, and conformal invariance
  in classical field theory.
\newblock {\em Journal of Mathematical Physics}, 52:012904, 2011.

\bibitem{BraDic61a}
C.H. Brans and R.H. Dicke.
\newblock Mach's principle and a relativistic theory of gravitation.
\newblock {\em Physical Review}, 124(3):925--935, 1961.

\bibitem{Fuj71a}
Y.~Fujii.
\newblock Dilaton and possible non-{N}ewtonian gravity.
\newblock {\em Nature Physical Science}, 234(9):5--7, 1971.

\bibitem{Fuj72a}
Y.~Fujii.
\newblock Scale invariance and gravity of hadrons.
\newblock {\em Annals of Physics}, 69:494--521, 1972.

\bibitem{FujMae04b}
Y.~Fujii and K.-I. Maeda.
\newblock {\em The Scalar-Tensor Theory of Gravitation}.
\newblock Cambridge University Press, 2004.

\bibitem{GasVen03a}
M.~Gasperini and G.~Veneziano.
\newblock The pre-big bang scenario in string cosmology.
\newblock {\em Physics Reports}, 373(1):1--212, 2003.

\bibitem{Bra05m}
C.H. Brans.
\newblock The roots of scalar-tensor theory: an approximate history.
\newblock 2005.
\newblock arXiv:[gr-qc]/0506063.

\bibitem{Wal84b}
R.~M. Wald.
\newblock {\em General Relativity}.
\newblock The University of Chicago Press, Chicago and London, 1984.

\bibitem{GalPas90b1}
A.~Galindo and P.~Pascual.
\newblock {\em Quantum Mechanics I.}
\newblock Springer Verlag, Berlin, etc., 1990.

\bibitem{Don11b}
Shi-Hai Dong.
\newblock {\em Wave Equations in Higher Dimensions}.
\newblock Springer, 2011.

\bibitem{GotEta92a}
M.~J. Gotay and J.~E. Marsden.
\newblock Stress-energy-momentum tensors and the {B}elinfante-{R}osenfeld
  formula.
\newblock {\em Contemporary Mathematics}, 132:367--392, 1992.

\bibitem{Dic62a}
R.~H. Dicke.
\newblock Mach's principle and invariance under transformation of units.
\newblock {\em Physical Review}, 125(6):2163, 1962.

\bibitem{Fla04a}
E.~E. Flanagan.
\newblock The conformal frame freedom in theories of gravitation.
\newblock {\em Classical and Quantum Gravity}, 21(15):3817, 2004.

\bibitem{FarNad07a}
V.~Faraoni and S.~Nadeau.
\newblock (pseudo) issue of the conformal frame revisited.
\newblock {\em Physical Review D}, 75(2):023501, 2007.

\bibitem{Car10b}
R~Carroll.
\newblock {\em On the emergence theme of physics}.
\newblock World Scientific, 2010.

\bibitem{Car10m}
R.~Carroll.
\newblock Remarks on gravity and quantum geometry.
\newblock arXiv:1007.4744[math.ph].

\bibitem{Koc10p}
B.~Koch.
\newblock Quantizing geometry or geometrizing the quantum?
\newblock In {\em QUANTUM THEORY: Reconsideration of Foundations - 5.}, volume
  1232 of {\em AIP Conference Proceedings}, pages 313--320. American Institute
  of Physics, 2010.
\newblock arXiv:1004.2879v2 [hep-th].

\bibitem{Koc09m1}
B.~Koch.
\newblock A geometrical dual to relativistic {B}ohmian mechanics - the multi
  particle case.
\newblock arXiv:0901.4106.

\bibitem{Smo06m}
L.~Smolin.
\newblock Could quantum mechanics be an approximation to another theory?
\newblock arXiv:quant-ph/0609109.

\bibitem{Schme11m}
I.~Schmelzer.
\newblock An answer to the {W}alstrom objection against {N}elsonian statistics.
\newblock arXiv:1101.5774[quant-ph].

\bibitem{CapLau11a}
S.~Capozziello and M.~De~Laurentis.
\newblock Extended theories of gravity.
\newblock {\em Physics Reports}, 509(4):167--321, 2011.

\end{thebibliography}

\end{document}